\documentclass{aa}  

\usepackage{graphicx}
\usepackage{txfonts}
\usepackage{lipsum}
\usepackage{subcaption}         
\usepackage{lscape}             
\usepackage{placeins}           
                                
\begin{document}

\title{Global statistical entropy and its implications for the main sequences of stars and galaxies}



   \author{D. Elbaz\inst{1}}

   \institute{
   Universit\'{e} Paris-Saclay, Université Paris Cit\'{e}, CEA, CNRS, AIM, 91191, Gif-sur-Yvette, France\\
             \email{david.elbaz@cea.fr}\\ }

   \date{Received June 30, 2025}

  \abstract
{In a dissipative system such as star or a galaxy, the emitted photons are decoupled from matter particles and may therefore be considered as part of a closed system to which the second law of thermodynamics applies. In the present work, I defined a global entropy using a statistical approach that accounts for the contributions of both matter particles and photons. The statistical contribution of radiation is described as a photon gas in the definition of this global entropy. 
The increase in global entropy can foster structure formation --rather than disorder-- because structures such as stars and galaxies are efficient in dissipating energy in the form of photons, and thus in producing entropy.
I show that stars generate a nearly equal amount of specific entropy and, therefore, a comparable number of photons per unit mass over their lifetime on the main sequence of the Hertzsprung--Russell (HR) diagram. This suggests that the main sequence of the HR diagram constitutes a locus of convergence toward a universal specific entropy production by stars. I then examined the implications of this approach for the star-formation main sequence in galaxies and found a similar result. 
The emergence of organized structures in cosmic history reflects the second law, as organized matter is efficient in generating entropy through the slicing of energy into lower frequency photons. This is also reflected in the dominant contribution of low-frequency photons to the extragalactic background light. Finally, in this paper I briefly discuss how this perspective may provide insight into the possibility of the existence of life elsewhere in the Universe.
}


   \maketitle
   \nolinenumbers

\section{Introduction}\label{sec1}
The first law of thermodynamics states that energy is conserved, but it says nothing about the direction of physical processes. That role is played by the second law, which governs the average direction of nonequilibrium dissipative phenomena. However, this law is probabilistic rather than deterministic: it only defines the most likely direction of evolution for an isolated system. Moreover, the notions of entropy production and dissipation are defined and used differently across the literature, which has led to significant ambiguity and potential confusion  \citep[see, e.g.,][]{dewar2014,lineweaver2014}.

Understanding the implications of the second law on the evolution of dissipative systems is of fundamental importance, since since complex structures only emerged in the history of the Universe once dissipation through radiation had become effective. Yet, this process is rarely addressed from a statistical perspective that explicitly accounts for the second law of thermodynamics. This is likely due to difficulties in defining the boundaries of a system. Indeed, once a system becomes dissipative, it is considered as an open system exchanging energy with its surroundings. Hence, it is no longer considered subject to the second law in a strict sense. 

The second law applies strictly to closed systems whose entropy can only increase or remain constant. The word entropy -coined by Clausius in 1865-- implicitly reflects this definition, as it combines the Greek words {trop\=e}, meaning transformation, and {in}, meaning within. Entropy thus describes the "transformations within," which highlights the fact that the second law only applies to an isolated system. The light that is radiated by a system is itself a source of entropy, but it is not usually considered as a part of the system. Instead, it is treated as energy leaving the system and being exchanged with the external environment. However, for an astrophysical source, such as a star or a galaxy, this radiation does not --at least on short timescales-- interact with the emitting system or with the outside world, which is essentially empty. The matter particles of such a system and the radiation they emit are effectively decoupled and may therefore be viewed as two components of a single closed system to which the second law can be applied.

This paper is structured as follows. I first define a global entropy using a statistical approach that accounts for the contributions of both matter particles and photons.
Radiation is described as a photon gas decoupled from matter and included as part of the system that was treated as closed, to which the second law of thermodynamics can be applied. 
I show that stars generate a nearly universal amount of entropy per unit mass over their lifetime on the Hertzsprung--Russell (HR) diagram. Interpreted within the framework of the fluctuation theorem, this result suggests that the main sequence (MS) of the HR diagram constitutes a locus of convergence toward a universal specific entropy production by stars. I then discuss the implications of this approach on the so-called star-formation main sequence (SFMS) of galaxies, which may similarly be interpreted as a locus of convergence in terms of specific entropy production by galaxies. Finally, I consider implications for the formation of structures, as traced by the cosmic background radiation, and in terms of the emergence of life in the Universe.

\section{Global entropy of dissipative systems and structure formation in the Universe}
\subsection{Definition of the global entropy}
The second law of thermodynamics states that the entropy of a closed isolated system must either stay constant or increase. If energy, in the form of heat, for example, is exchanged with the outside world, the system is considered as open. The dissipation of energy is generally described by the term on the right of the inequality in Eq.~(\ref{EQ:DQ}), below, where $\Delta$$Q$ is the heat exchanged with the outside (this inequality becomes an equality in the case of a reversible heat exchange):
\begin{equation}
\Delta S \geq \Delta Q / T 
\label{EQ:DQ}
.\end{equation}
This definition of entropy and of the second law of thermodynamics, depends on what is considered to be part of the system or of the outside world. In the case of dissipative structures releasing light into the essentially empty space of the Universe, the light released in the process does not interact with the system itself. In addition, when it interacts with the outside world, the system is not affected either. As a first approximation, we can therefore consider that in the time following the emission of this light, it behaves as a photon gas that can be considered part of the system. This could be the case of a star radiating its photons into interstellar space or a galaxy releasing its radiation into intergalactic space. Here, I define the global entropy, $S_{global}$, as the sum of the statistical entropies associated with the matter particles and the photon gas (Eq.~\ref{EQ:S0global}):
\begin{equation}
S_{global} = S_{part} + S_{photons} 
\label{EQ:S0global}
.\end{equation}

Boltzmann's definition of the statistical entropy of a collection of $N_{part}$ matter particles is given in Eq.~(\ref{EQ:Boltzmann}), where $\Omega$ measures the number of microscopic states energetically available to the system (positions and velocities of the particles), and k$_B$ is the Boltzmann constant (in J K$^{-1}$): 
 
\begin{equation}
S_{part} =~k_{B}~\log \Omega
\label{EQ:Boltzmann}
.\end{equation}If we consider a star free-floating in empty space, the outside world is the vacuum that surrounds it, and the heat it exchanges with the surrounding vacuum is strictly in the form of photons. Once emitted, the photons are decoupled from the star and travel freely in the vacuum. The resulting radiation can thus be described as a photon gas that is part of a system that may be considered closed and isolated. The system is indeed globally isolated from the outside world, even if the photon gas extends out into space. The second law of thermodynamics can therefore be applied, provided that the entropy of the system includes the contributions of both matter and light particles. In its general form, the entropy of a photon gas, $S_{phot}^{gen}$, is given by Eq.~(\ref{EQ:S0phot}): 
\begin{equation}
S_{phot}^{gen} = -\frac{2 k_B}{c^3} \int_{\nu=0}^{\infty} \int \nu^2 \left[ n \log n - (n+1) \log(n+1) \right] \text{d}\Omega_{solid} \text{d}\nu
\label{EQ:S0phot}
,\end{equation}
where d$\Omega_{solid}$
is the elementary solid angle around the direction of propagation of the
photons, $\nu$ is their frequency, and $n$ is the photon occupation number
at the frequency $\nu$. In the present paper, I use a simplified version of this formula in the case of the black-body radiation of a system at thermodynamic equilibrium, such as a star (or a collection of stars in a galaxy), which is given in Eq.~(\ref{EQ:Sphotons}) \citep[see Appendix~\ref{secA} and, e.g.,][]{Kelly1981,Reif1965}:
\begin{equation}
S_{photons} =~3.6~k_{B}~N_{photons}
\label{EQ:Sphotons}
.\end{equation}Hence, it is as if each photon were carrying a unit quantity of entropy of 3.6 in units of $k_B$.
In the absence of interactions between the emitted photons and matter, the two components of entropy can be added together, leading to the definition of a global entropy, $S_{global}$ (Eq.~\ref{EQ:Sglobal}). This amounts to transferring the term on the right of Eq.~(\ref{EQ:DQ}) --associated with the  exchange of heat with the outside world-- to the term on the left --which defines the entropy of the system made up of matter particles and photons: 
\begin{equation}
S_{global} = S_{part} + S_{photons} = k_{B}\log \Omega + 3.6~k_{B} N_{photons}
\label{EQ:Sglobal}
.\end{equation}

In the classical perspective where the system is described as open, entropy can only increase with increasing disorder. If we instead consider the global entropy, which includes the statistical contribution of the photon component, then an increase in entropy can either be due to an increase of disorder or order if it is accompanied with an even more efficient production of entropy in the form of photons (since the photon gas belongs to the system considered as a closed one). 
The statistical definition of the global entropy (Eq.~\ref{EQ:Sglobal}) includes the number of photons, but not their frequency. This suggests that if an ensemble of matter particles evolve toward a state in which dissipation happens in a way that produces a large number of photons, this state will generate an increase of global entropy, despite it being "ordered."

\subsection{The fluctuation theorem}
\label{SEC:fluct}
The fluctuation theorem was introduced by \cite{evans1994} as a generalization of the second law of thermodynamics applied to small systems, far from equilibrium. As summarized in their later review \citep{evans2002}, this relation was applied to the special case of dissipative nonequilibrium systems subject to constant energy "thermostatting." When applied to a system on a nanoscale, it explains how ergodic movements of particles can lead to a decrease in entropy for a short time, but globally evolve in a direction that systematically increases entropy. This statistical approach thus makes it possible to explain, from a probabilistic mathematical formulation, how a microphysics that is subject to pure reversibility can lead to irreversibility (i.e., a direction of increasing entropy), hence solving the so-called Loschmidt paradox \citep{loschmidt1876} regarding the enigmatic origin of irreversibility in  a world governed by physical laws that are reversible in time. For a full account of the various formulations of the fluctuation theorem, I refer the reader to the review by \cite{sevick2008}, where the authors discuss how "fluctuation theorems have resulted in fundamental breakthroughs in our understanding of how irreversibility emerges from reversible dynamics."

When integrated over long timescales and macroscopic scales, the fluctuation theorem recovers the second law since small-scale ergodic movements do not lead to macroscopic and long-term paths in which entropy is reduced. However, it also adds a probabilistic viewpoint that can be used to quantify how different paths --which all follow the second law by increasing entropy-- compare to each other; that is, namely, to decipher which one is more probable than the others. While Clausius’s statement of the second law asserts that entropy tends to an increase, the fluctuation theorem shows that the probability of such evolution increases with the magnitude of the entropy production. So, while the second law indiscriminately favors all paths that increase the entropy of a closed system, the fluctuation theorem attributes a higher probability to paths that lead to greater entropy. This theorem can be seen as an incarnation of the idea that a "macrostate of higher entropy can be realized in overwhelmingly more ways" and is therefore more probable \citep{jaynes1985}.

The fluctuation theorem tells us that the ratio of the probability, $P_2$, that a system evolves in a forward direction from state 1 to state 2 over the probability, $P_1$, for the reverse direction, is given by the exponential of the difference of entropy, $\Delta S=S_2-S_1$. In the present study, I used the formulation of the fluctuation theorem given by Eq.~(\ref{EQ:expDS}) \citep{gaspard2011}:
\begin{equation}
\frac{P_2}{P_1} = \exp(\Delta S/k_{B}) = \exp\left[(S_2 - S_1)/k_{B}\right]
\label{EQ:expDS}
.\end{equation}
This indicates that a system has an exponentially higher probability of evolving toward a state of entropy, $S_2$, rather than one associated with $S_1$, with the probability ratio between the two states given by Eq.~(\ref{EQ:expDS}). 

\section{Implications of the global entropy and second law on cosmic structures}
\subsection{Universality of the global entropy produced by stars during their lifetime on the MS}
\label{SEC:HR}
Stars spend most of their lives in the MS of the HR diagram, showing a correlation between their luminosity and effective temperature \citep[see, e.g.,][]{gaia2018}. During this period, they maintain the same size and effective surface temperature until they have completed their hydrogen-core fusion and begun the fusion of helium. A star on the MS maintains hydrostatic equilibrium, its gravitational collapse being balanced by internal pressure, which is dominated by thermal gas pressure in low-to-intermediate-mass stars and by radiation pressure in high-mass stars.

To a first approximation, the global entropy of a star on the MS varies due to nuclear fusion in the core ($\Delta S_{fusion}$) and the emission of photons ($\Delta S_{photon}$). It can be seen that $\Delta S_{fusion}$ is completely negligible compared to $\Delta S_{photon}$ by a factor of about one million (see Appendix~\ref{secB}). We can therefore neglect the entropy change associated with nuclear fusion; since to a first approximation the size and temperature of the star do not vary as long as it remains on the MS, we can assume that the main driver of the entropy of a star is given by Eq.~(\ref{EQ:Sphotons}). Hence, although it offers a smaller number of microstates to a collection of matter particles, a star represents a state of matter that follows the second law by increasing entropy mainly through radiation. However, stars of different masses do so with different levels of efficiency. A massive star generates more photons per unit mass and time and, hence, more entropy than a lower mass star, but it does so over a shorter lifetime. I now propose to compute the integrated entropy generated by stars while on the MS.

\begin{figure}[ht]
  \begin{center}
 \includegraphics[width=7.5cm]{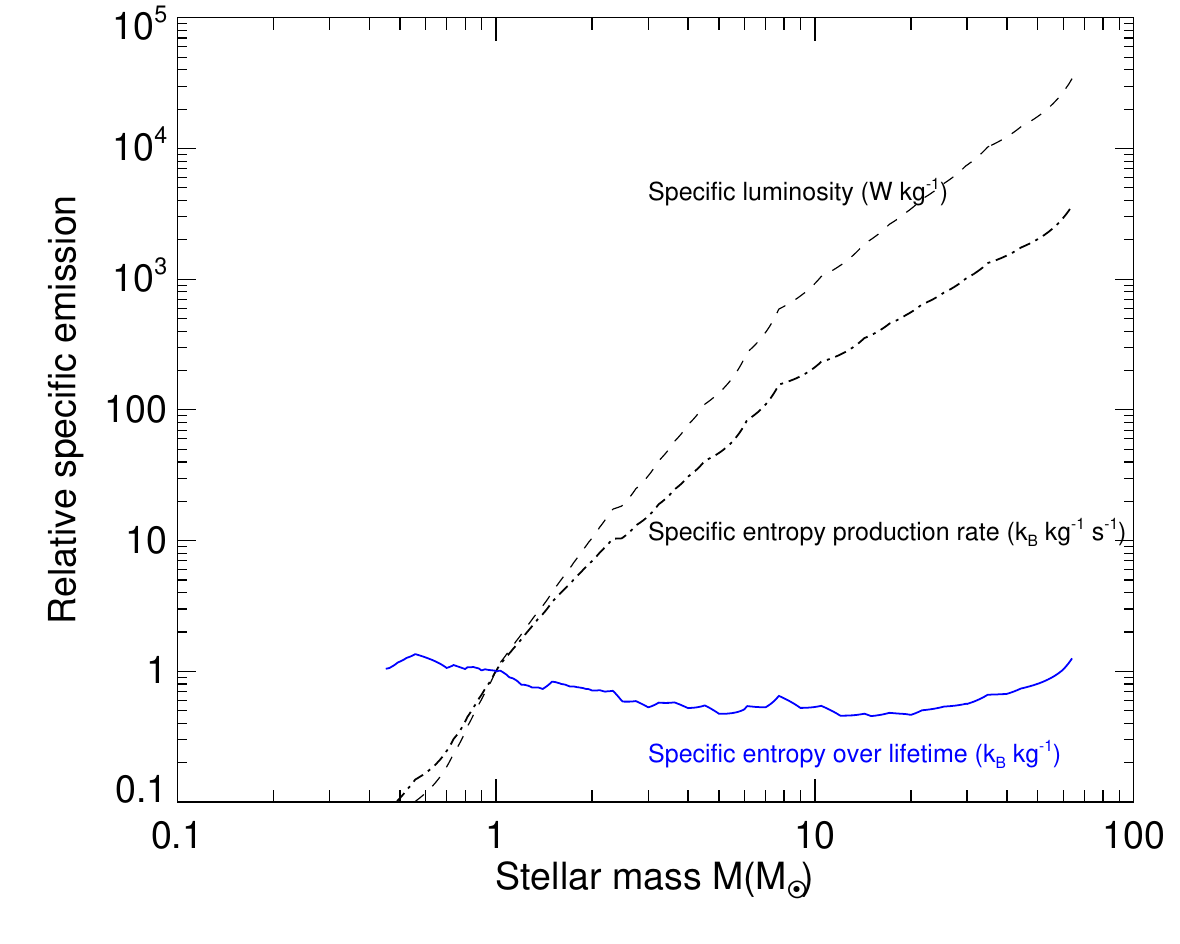}
 \caption{Comparison of specific luminosity (dashed line), specific entropy production (dash-dotted line), and specific entropy production integrated over the lifetime of a star on the MS of the HR diagram (solid blue line) for stars of different masses. All curves are normalized to a value of one at the mass of the Sun.
}
 \label{FIG:S_MS}
  \end{center}
 \end{figure}

The luminosity of stars varies proportionally to the stellar mass to the power of $\alpha$ ($L\sim M^{\alpha}$) greater than one; hence, the specific luminosity of a star ($L/M$ in W kg$^{-1}$) increases with stellar mass (see the dashed line in Fig. \ref{FIG:S_MS}). The change in the slope, $\alpha,$ with stellar mass comes from the internal physics of stars, and corresponds to transitions between regimes in the stellar interior physics, such as the transition between regimes where the star is supported against gravitational collapse by thermal or radiation pressure. Here, I used the values from Table 4 in \cite{eker2018}, where $\alpha$=2.028 at the lowest masses (0.179--0.45 $M_{\odot}$) and increases to $\alpha$$\sim$ 4 -- 5 at masses between 0.45 and 7 $M_{\odot}$ ($\alpha$=4.572 at 0.45--0.72 $M_{\odot}$, 5.743 at 0.72--1.05 $M_{\odot}$, 4.329 at 1.05--2.40 $M_{\odot}$, 4.329 at 1.05--2.40 $M_{\odot}$, and 3.967 at 2.4--7 $M_{\odot}$) before decreasing to $\alpha$= 2.865 at the highest masses (7--31 $M_{\odot}$).

Since the entropy production of a star is directly proportional to the number of photons it generates, we can derive the specific entropy production rate of a star on the MS (i.e., the amount of entropy produced through photon production per unit mass of the star) from its luminosity and radiation frequency. Here, I assume that the photons are radiated at the peak wavelength of the black body given by Wien's law. Therefore, I simply assumed the difference in wavelength between stars of different masses to be inversely proportional to the effective temperature. The resulting specific entropy production rate is shown with a dash-dotted line in Fig. \ref{FIG:S_MS}. The slope is less steep than for the specific luminosity of stars because stars of higher masses present a higher surface temperature; hence, they generate fewer photon-per-unit energy \citep[I used the Table 7 in][]{eker2018}.

The specific amount of entropy generated by a star during its lifetime on the MS ($\tau_{MS}$) is given by Eq.~(\ref{EQ:Nphstar}), and the global entropy of the star is given by Eq.~(\ref{EQ:Sg1}):
\begin{equation}
S_{photons}^{MS, spec} \propto N_{photons}^{MS}/M_{\star} \propto (L_{MS}/M_{\star}) \times \tau_{MS} \times \lambda_{photons}~[k_B~kg^{-1}]
\label{EQ:Nphstar}
,\end{equation}
\begin{equation}
S_{global} = k_{B}\log \Omega + 3.6~k_{B} N_{photons}^{MS} \simeq 3.6~k_{B} N_{photons}^{MS}~[k_B]
\label{EQ:Sg1}
.\end{equation}

As discussed above, the entropy reduction due to nuclear fusion is negligible (see Appendix~\ref{secB}). During its lifetime on the MS \citep[$\tau_{MS}$ from Table 6 of][]{bressan1993}, the global entropy of a star, which is driven by the emission of photons, is given by the solid blue line in Fig. \ref{FIG:S_MS}. If we compare stars of all masses in the MS, we find that they share an interesting common property: all stars generate nearly the same integrated global entropy per unit mass over their lifetime on the MS, independently of their mass. Hence, the HR diagram is a place of convergence for collections of particles to generate entropy with a universal efficiency. We did not extend this analysis to the following phases of stars, such as the red or blue giant branch, among others.

The first-order scaling laws generally assumed for stars on the MS would give us a trend that is exactly flat.
In a first approximation, stars on the MS radiate a luminosity, $L_{MS}$, that is proportional to $M_{\star}^{3.5}$; hence, $L_{MS}/M_{\star}$$\propto$$M_{\star}^{2.5}$. The lifespan of a star on the MS is on the order of $\tau_{MS}$$\propto$$M_{\star}^{-2}$. The number of photons that are radiated depends on their wavelength, which is inversely proportional to the surface temperature of the star, $T_{MS}$. Since $T_{MS}$ is related to the mass as $T_{MS}$$\propto$$M_{\star}^{0.5}$, it results that $\lambda_{photons}$$\propto$$M_{\star}^{-0.5}$. Here again, we find that the number of photons radiated per unit mass and, hence, the specific entropy produced by a star during its lifetime on the MS is nearly independent of the stellar mass (see Eq.~\ref{EQ:N2phstar}):
\begin{equation}
N_{photons}^{MS} \propto (L_{MS}/M_{\star}) \times \tau_{MS} \times \lambda_{photons} \propto M_{\star}^{2.5} \times M_{\star}^{-2} \times M_{\star}^{-0.5}   
\label{EQ:N2phstar}
.\end{equation}

In conclusion, more massive stars are more efficient in producing entropy per unit mass and time, but they emit fewer photons per unit energy and spend less time doing so on the MS. As a result, if we compare the state of a star with its surrounding photon gas, as measured by its global entropy integrated over the star's lifetime on the MS, stars of different masses generate an almost equal amount of specific entropy (solid blue curve in Fig. \ref{FIG:S_MS}). The fluctuation theorem (see Sect.~\ref{SEC:fluct}) tells us that the probability of a system evolving from state 1 to state 2 is given by the exponential of the difference in entropy between the two states (Eq.\ref{EQ:expDS}). Therefore, the fact that stars of a wide range of masses end up producing almost the same amount of entropy per unit mass throughout their lifetime on the MS of the HR diagram suggests that they are evolving toward a universal specific entropy. The MS of the HR diagram can therefore be considered as a locus of convergence toward a universal specific entropy production of matter by stars.

\subsection{Universality of the global entropy produced by star-forming galaxies on the SFMS}
\label{SEC:SFMS}
A universal relation connects the integrated luminosity on a galaxy scale with its stellar mass. This relation is reminiscent of the stellar MS, but it concerns all the stars of a galaxy, hence its name, the SFMS (see \citealt{elbaz2007}, \citealt{daddi2007}, \citealt{noeske2007}, \citealt{schreiber2015}, and \citealt{popesso2023} and references therein). This relation only applies to star-forming galaxies; it does not apply to passive galaxies. The luminosity is therefore directly related to the star-formation rate (SFR) of galaxies. It is thus generally represented as a correlation between the SFR and the stellar mass of galaxies, $M_{\star}$. It takes the form of a power law with a slope close to unity, with a value on the order of $\alpha = 0.9$ for
SFR = $A(z)\times$ M$_{\star}^{\alpha}$, where $A(z)$ is a normalization factor that depends on redshift.

Although it is natural for more massive galaxies to emit more light, the SFMS has two striking features. First, the SFR--M$_{\star}$ correlation displays a remarkably small scatter of only about 0.3 dex. This indicates that galaxies form most of their stars on this sequence and that events such as starbursts, during which a galaxy forms stars at a rate of a factor of more than two to three above the SFMS median, contribute only marginally to the buildup of the stellar mass of present-day galaxies \citep[see, e.g.,][]{rodighiero2011}. Second, the slope of the SFMS is close to one, which means that galaxies form stars with nearly the same specific SFR (sSFR=SFR/M$_{\star}$; i.e., star formation per unit stellar mass) independently of their total stellar mass. This has been observed to take place over the bulk of cosmic history from the local $z\sim0$ Universe to at least $z$$\sim$5--6, although with an evolving normalization (see, e.g., \citealt{schreiber2017} and \citealt{ciesla2024} and references therein). The normalization of the SFMS, $A(z)$, evolves with cosmic time, but the shape of the SFMS remains nearly the same at all cosmic epochs with a slope close to unity. Hence, the sSFR is universal and independent of stellar mass at each cosmic epoch.

Moreover, it has been found that this universal relation is accompanied by a universality of the dust temperature of galaxies at any given redshift \citep[see, e.g.,][]{magnelli2014}. Most of the luminosity generated by star formation is radiated in the UV and absorbed by interstellar dust, which reemits it in the far-infrared. Since the dust temperature of galaxies on the SFMS is independent of galaxy mass, this implies that the number of emitted photons does not depend on mass either. This in turn implies that the specific entropy produced by galaxies on the SFMS --thus, during most of a galaxy’s lifetime, since the majority of stars are formed while galaxies lie on the SFMS-- is also universal, similarly to that produced by stars on the MS of the HR diagram.

\begin{figure}[ht]
  \begin{center}
   \includegraphics[width=7.5cm]{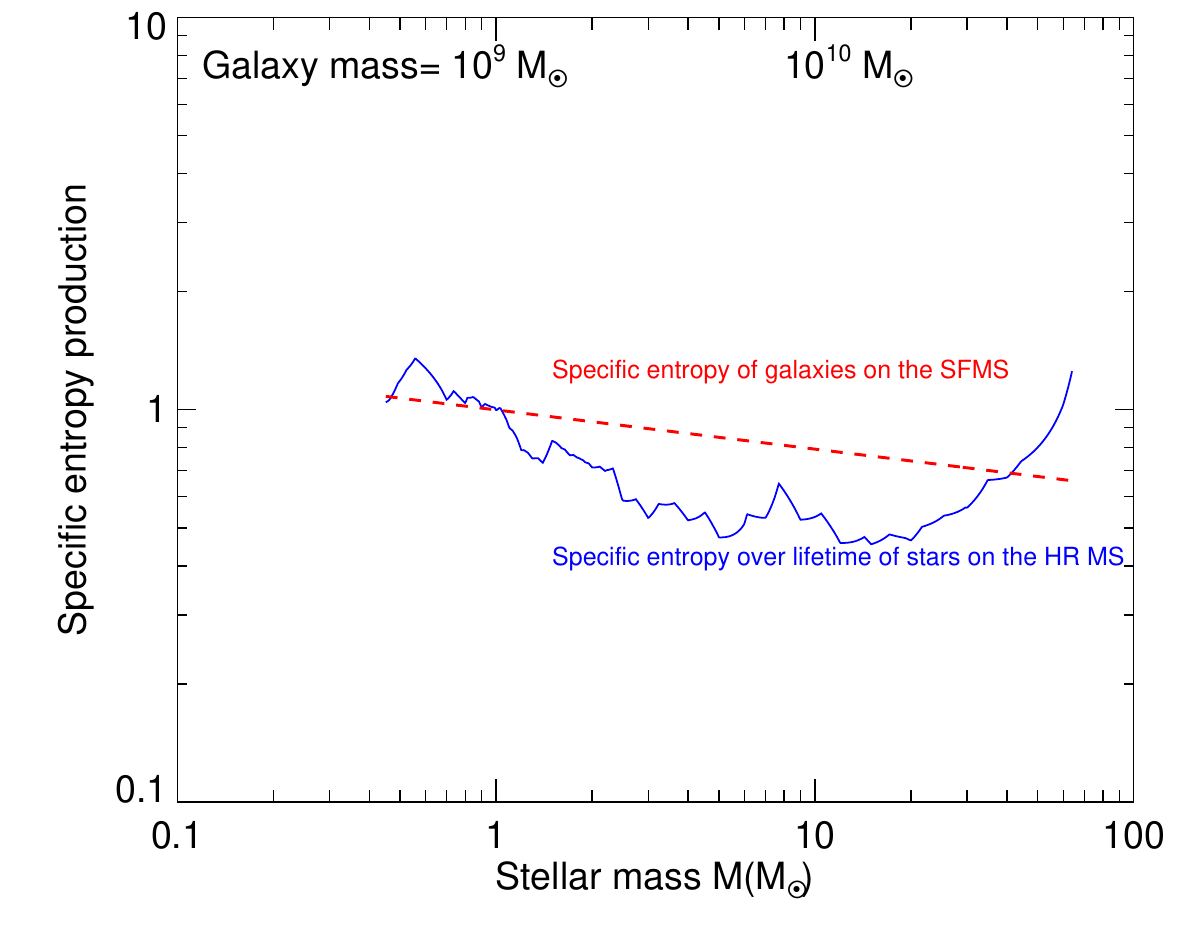}
 \caption{Comparison of specific entropy production integrated over the lifetime of a star on the MS of the HR diagram (solid blue line) for stars of different masses (normalized to unity for a solar-mass star) with the specific entropy production of galaxies on the SFMS for galaxies of different masses (dashed red line; upper horizontal axis). The normalization of the SFMS evolves with redshift; here, it is normalized to unity at 10$^9$ M$_{\odot}$.}
 \label{FIG:Gal_MS}
  \end{center}
 \end{figure}

This means that galaxies form the bulk of their stars in a self-regulated mode reminiscent of the stars on the HR MS diagram. I compare both trends in Fig.\ref{FIG:Gal_MS}, where galaxies are shown with a dashed red line (normalized to unity for a stellar mass of 10$^9$ M$_{\odot}$) and stars with a solid blue line. It is believed that enhanced star formation in galaxies is associated with a higher supernova rate, which in turn decreases the gas density within the galaxy and reduces the SFR. Since the supernova rate drops together with the decrease in star-formation, the mechanism becomes self-regulated. We note that the MS shows a bending at high masses below $z$$\sim$3-4, which is interpreted as a sign of a transition from the cold to the hot accretion regime \citep{daddi2022}.

In contrast to stars on the MS of the HR diagram, galaxies on the SFMS undergo mass evolution (i.e., mass growth), yet they remain on the SFMS for most of their lifetime. Given the observation that all galaxies generate a roughly equal amount of specific entropy per unit mass, and considering that the same population of galaxies progresses from the low-mass to the high-mass end of the SFMS over cosmic time, this suggests that SFMS galaxies ultimately produce comparable total amounts of entropy per unit mass over their entire lifetime. Hence, from a statistical perspective inferred from the fluctuation theorem, the SFMS may be viewed as a locus of convergence toward a universal specific entropy production of matter by galaxies. 

\subsection{Global entropy and the cosmic background}
The second law is of course probabilistic rather than deterministic; hence, we cannot deterministically infer which path will be followed by an ensemble of particles on the basis of the second law alone. However, there is a general trend in the history of structure formation in the Universe, in which matter becomes organized into structures that are increasingly efficient at dissipating energy into a growing number of photons. This is a natural consequence of the second law and of the fluctuation theorem, since entropy is directly proportional to the number of photons radiated by a dissipative system.

Starting from the ionized intergalactic matter following reionization, matter evolved from the simplest and least structured systems to the densest and most organized ones through a succession of radiative mechanisms that come into play as density increases. I briefly describe these mechanisms below. At the lowest densities, cooling of the hot, ionized intergalactic medium, at temperatures of 10$^6$ to 10$^8$ K (1 to 10 keV), produces photons at wavelengths between 0.1 and 10 nm. At higher densities, UV radiation dominates through collisional excitation of C, O, and Si ions (at 100-300 nm) down to temperatures of about 10$^5$ K, followed by Ly$\alpha$ emission from hydrogen at 121.6 nm down to $\sim$10$^4$ K. Below this transition temperature, further radiative energy loss becomes less efficient, as matter becomes neutral with the formation of atoms and, in the near-absence of free electrons, collisions are no longer sufficient to excite electronic levels. Cooling then proceeds through molecular emission, in particular the rotational transitions of CO molecules, excited by collisions with hydrogen molecules (whose dipole is too weak to produce significant radiation). These transitions range from the high J=6-5 transition at 0.43 mm (691 GHz) to the fundamental J=1-0 line at 2.6 mm (115 GHz). At even higher densities, dust emission becomes dominant, radiating primarily in the far-infrared ($\sim$100$\mu$m) and millimeter ($\sim$1 mm) regimes. These processes also correspond to an increasing efficiency in entropy production per unit of energy released, as the emitted photons have a longer wavelength and therefore carry less energy per photon. Structure formation is thus associated with an increase in the entropy of the Universe. 

\begin{figure}[ht]
  \begin{center}
   \includegraphics[width=7.5cm]{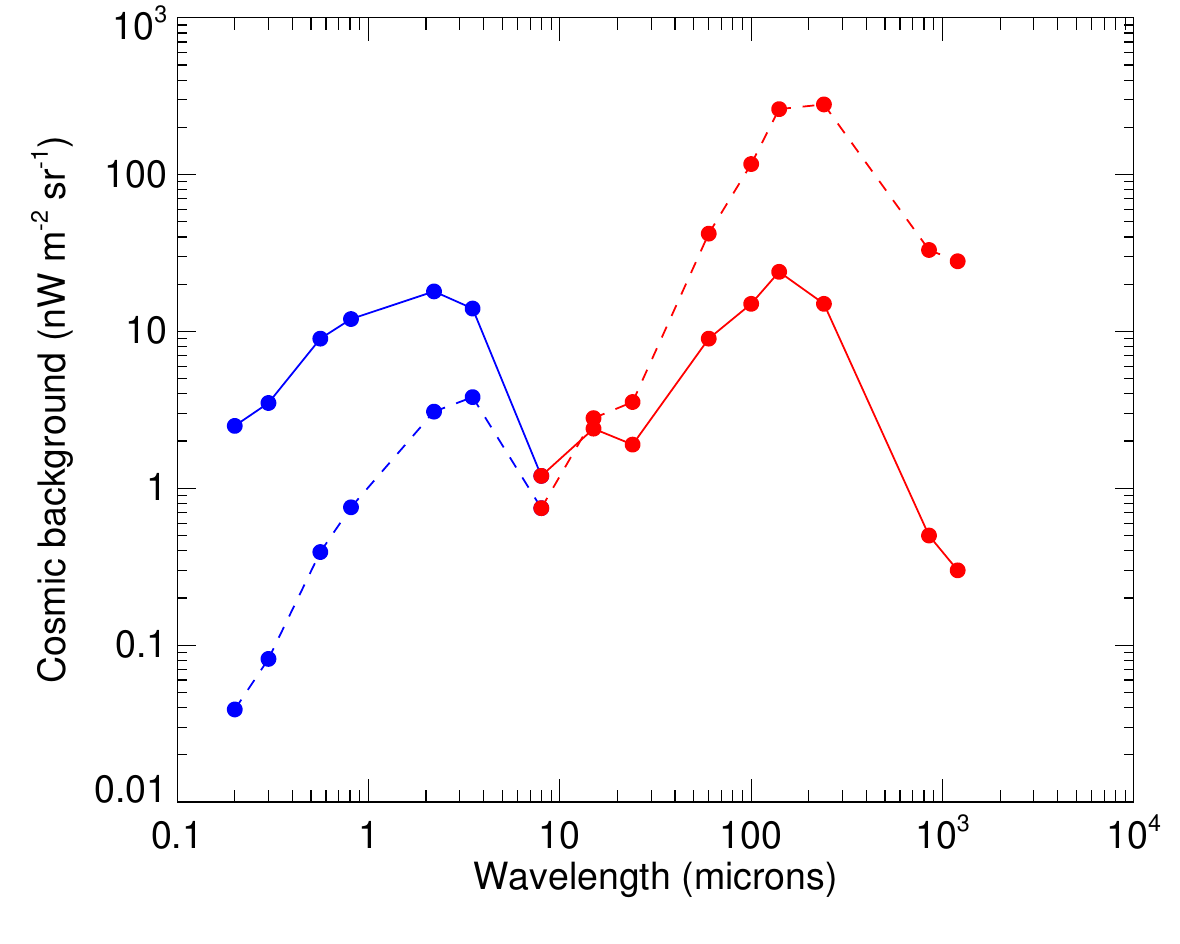}
 \caption{Extragalactic background light generated by star formation in galaxies since cosmic dawn (solid lines), shown in blue on the optical side and as a red line in the infrared, where stellar light is reprocessed by dust (data from \citealt{lagache2005}). The dashed lines show the relative amount of entropy, traced by the number of photons, that has accumulated and been carried by photons throughout the history of star formation in galaxies.}
 \label{FIG:CB}
  \end{center}
 \end{figure}

The efficiency of the processes that convert the energy associated with structure formation into an increasing number of photons and, hence, entropy are directly visible in the diagram showing the relative contributions of the cosmic background light from the UV to the millimeter range \citep[see Fig. \ref{FIG:CB}, data in nW m$^{-2}$ sr$^{-1}$ from][]{lagache2005}. It has been shown that the power radiated during structure formation over the entire cosmic history is nearly equally divided between the optical wavelength range, that is, the cosmic optical background (COB; solid blue line; COB= 23 nW m$^{-2}$ sr$^{-1})$  and the infrared range, that is, the cosmic infrared background (CIB; solid red line; CIB= 24 nW m$^{-2}$ sr$^{-1}$) \citep{dole2006}. However, the number of photons produced, and thus the entropy generated, depends on the energy carried by individual photons and, therefore, on their wavelength. This is shown with the dashed lines in Fig. \ref{FIG:CB} (normalized to one at a wavelength of one micron). It follows that the global entropy associated with the CIB --shown with a dashed red line, $S_{photons}^{CIB}$ $=$ 2.7 $\times$ 10$^{86}$ k$_B$-- is about 80 times greater than that of the COB, which is shown with a dashed blue line: $S_{photons}^{COB}$ $=$ 3.4 $\times$ 10$^{84}$ k$_B$.

The dominant contribution of long wavelengths to the cosmic-background energy budget illustrates the conversion of a given amount of energy into a larger number of photons and, thus, the efficient production of entropy, which plays a major role in the history of structure formation in the Universe. This suggests that structure formation is favored in the Universe, not only because it is a natural consequence of gravitation, but also an efficient pathway toward a global increase of entropy. Structure formation is therefore associated with an increasingly efficient slicing of energy into photons of lower energy and, hence, by a progressively more efficient production of entropy. Interestingly, one of the main quantitative differences --possibly the dominant one-- among different epochs of cosmic history lies in the total number of photons present in the Universe (while cosmic microwave background photons are more numerous, their number remains essentially constant). It is important to recall, however, that these considerations are probabilistic rather than deterministic. This implies that neither the exact evolutionary path nor the timescale over which it unfolds is imposed by the second law. The preceding considerations should therefore be viewed as an a posteriori confirmation of a generic process at work, rather than as a predictive framework for determining which path matter will follow or on what timescale.

This budget also includes the formation of black holes, since they form through dissipative processes (excluding the possible existence of primordial black holes). If black holes are indeed destined to evaporate into photons \citep{hawking1974}, this trend will become even more pronounced in the far future. However, black-hole evaporation is an extremely slow process. 

\subsection{The second law and the emergence of life}
In his book What is life?, \cite{schrodinger1944} investigated the apparent contradiction between the emergence of life and the second law of thermodynamics for the first
time. Schr\"odinger showed that this conundrum can be resolved if life is considered as an open system. As stars, animals may be seen as systems made of matter particles and surrounded by a gas of photons that they emit. 

If we restrict ourselves to the specific entropy generated by a star such as the Sun and compare it to that generated by an adult human being, it can be shown (see Appendix \ref{secC}) that a human being radiates 200,000 more photons per unit mass and time than the Sun. A human being is therefore also about 200,000 times more efficient in producing entropy per unit mass and time than the Sun. Since systems that are more efficient in producing entropy are associated with an exponentially higher probability of forming, this may be seen as a good reason to consider that life could exist beyond our planet. 

From this purely thermodynamic perspective, life may be interpreted as a highly probable outcome of the evolution of matter in the Universe, possibly even more probable than the formation of a star. However, I emphasize that this probabilistic argument does not demonstrate that life necessarily exists elsewhere in the Universe. Rather, it suggests that the emergence of life is not in contradiction with the global thermodynamic evolution of matter and structure formation in the Universe. It does, however, provide a strong motivation to search for the existence of life on exoplanets.

\section{Conclusion}
In this paper, I propose including radiation in the definition of entropy, using a statistical formulation in which entropy is directly proportional to the number of photons. This global statistical entropy therefore includes a component for matter and another associated with radiation --treated as a photon gas surrounding the system-- which can be summed since they are decoupled. 

Stars spend most of their lives on the MS of the HR diagram, during which they convert hydrogen into helium. I show that, when considering a star together with its surrounding photon gas and measuring the global entropy integrated over its MS lifetime, stars of different masses generate nearly the same specific entropy, in the form of photons, per unit mass over their lifetime. Although more massive stars produce entropy more efficiently per unit mass and time, they radiate fewer photons per unit energy and spend less time on the MS. This suggests that the MS of the HR diagram constitutes a locus of convergence toward a universal specific entropy production by stars.

In an analogous manner, but on larger scales, galaxies form most of their stars in a secular star-formation mode known as the SFMS, which is reminiscent of the stellar MS in that both galaxies and stars spend most of their respective MS lifetimes there. In both cases, the MS is thought to result from self-regulation, that is, a balance between gravity and the energy released through nuclear fusion, within stars, and supernova explosions, within galaxies. We find that galaxies on the SFMS also produce a specific entropy that varies only weakly with galaxy mass, in a way analogous to what is observed for individual stars. This suggests that the SFMS also constitutes a locus of convergence toward a universal specific entropy production by galaxies.

We then considered the global entropy production associated with structure formation over cosmic history, as traced by the cosmic optical and infrared backgrounds (i.e., non-CMB photons). This comparison shows that dust reprocessing of stellar light is a highly favored process since it grinds energy into a larger number of photons and therefore produces entropy more efficiently -- a tendency favored by the fluctuation theorem. Finally, we discussed how the fact that living systems are more than stars at efficient in producing entropy per unit mass may be interpreted as a motivation to consider that life could exist elsewhere in the universe.

The consistent statistical treatment of matter and energy was probably what Albert Einstein had in mind when he wrote the following in 1905: "I shall show in a separate paper that when considering thermal phenomena, it is completely sufficient to use the so-called `statistical probability,' and I hope thus to do away with a logical difficulty which is hampering the consistent application of Boltzmann’s principle”  \citep{einstein1905}. This “separate paper” was unfortunately never published.

\begin{acknowledgements}
The author would like to thank for their precious support: 
Michel Cassé,
Catherine Cesarsky, 
Emanuele Daddi,
Mauro Giavalisco, 
Patrick Hennebelle, 
Kirone Mallick, 
Joel Primack, 
Pascal Tremblin. 
\end{acknowledgements}

%

\bibliographystyle{aa}
\bibliography{biblio_entropy}

\begin{appendix}




\onecolumn
\section{Entropy of a gas of photons}\label{secA}
%
If a system made of $N_{part}$ particles, within a fixed volume V, is at thermodynamic equilibrium at a temperature T, its entropy, S, and the gas of photons radiated by the system containing $N_{photons}$ are defined by Eqs.~\ref{EQ:ST} and~\ref{EQ:Nph}, respectively \citep[see, e.g.,][]{Kelly1981,Reif1965}, with $\zeta(3)$=1.202 and $b=8\pi^{5}k_{B}^{4}/(15h^{3}c^{3})$:
\begin{equation}
S_{photons} = \frac{4}{3}~b~V~T^3 
\label{EQ:ST}
\end{equation}
\begin{equation}
N_{photons} = \left[ \frac{30~\zeta(3)}{\pi^4 k_B} \right]~b~V~T^3 
\label{EQ:Nph}
.\end{equation}Hence, we can derive the entropy of a gas of photons (Eq.~\ref{EQ:Sph}):
\begin{equation}
S_{photons} = \left[ \frac{4~\pi^4}{90~\zeta(3)} \right] N_{photons} k_B = 3.6~N_{photons} k_B
\label{EQ:Sph}
.\end{equation}

\section{Global entropy produced by a star}\label{secB}
In a given duration, e.g. one second, a star of mass M converts a fraction, f, of its mass of hydrogen into helium. This leads to a reduction of its entropy due to a diminution of these f particles by a factor four, and the associated reduction of number of microstates. If we assume that the star is a perfect gas, the entropy of a perfect gas of N particles in a volume V, at constant temperature, can be approximated by the Sackur-Tetrode equation \citep[see, e.g.,][]{binder2023}:

\begin{equation}
S = N k_B \left[ \ln\left( \frac{V}{N \Lambda^3} \right) + \frac{5}{2} \right] 
.\end{equation}

Since our goal is to determine the variation of entropy after the fusion of a fraction (f) of the mass of hydrogen, the $\Lambda$ term can be taken out of the logarithm as a constant additive term:

\begin{equation}
\frac{S}{N k_B} = - ln N + C 
,\end{equation}\noindent where
\begin{equation}
C = \ln\left( \frac{V}{\Lambda^3} \right) + \frac{5}{2} 
,\end{equation}and $\Lambda$ is the thermal de Broglie wavelength:

\begin{equation}
\Lambda = h / \sqrt{2 \pi m k_B T} 
.\end{equation}

\noindent This term depends on the temperature of the system, which we consider here as constant, as well as the occupied volume, for a star on the MS. In the case of the Sun ($T_{\odot} = 5770$ K, $V_{\odot} = 1.41 \times 10^{27}$ m$^3$, hence C = 138.5), every second, a tiny fraction of the Sun's mass is involved in nuclear fusion (6$\times$10$^{11}$ kg to be compared to the mass of the Sun of 2$\times$10$^{30}$ kg). Hence, a very small mass fraction of $f$ (=3$\times$10$^{-19}$) gets a reduction in the number of microstates by a factor of four. It results that the number of particles after the fusion becomes
\begin{equation}
N' = (1 - f)N + \frac{fN}{4} = N \left( 1 - \frac{3f}{4} \right)
.\end{equation}And the entropy, $S'$, for these $N'$ particles becomes
\begin{equation}
S' = N' k_B \left[ \ln\left( \frac{V}{N' \Lambda^3} \right) + \frac{5}{2} \right]
.\end{equation}After replacing N' with its value in the definition of S', we find the following value for $\Delta S$=$S'-S$ per particle and per unit $k_B$:

\begin{equation}
\frac{\Delta S}{N k_B} = \frac{3f}{4} \left[ ln N + \ln\left( \frac{V}{\Lambda^3} \right) - \frac{3}{2} \right] 
.\end{equation}

In the case of the Sun ($N_{\odot}=1.2 \times 10^{57}$ particles), this gives
\begin{equation}
\frac{\Delta S}{N k_B} = -1.35 \times 10^{-18}
.\end{equation}
Hence, the loss of entropy (per unit time, s) by matter particles due to fusion in the Sun is
\begin{equation}
\Delta S_{fusion} = -1.6 \times 10^{39} k_B
,\end{equation}
which must be compared with the entropy increase $\Delta S_{photon}$:
\begin{equation}
\Delta S_{photon} = 3.6~k_{B}~N_{photons}
.\end{equation}

Every second, a star like the Sun emits $L_{\odot}$=3.826$\times$10$^{26}$ W of photons at a typical wavelength of 0.5\,$\mu$m, or $\sim$10$^{45}$ photons. Hence,
\begin{equation}
\Delta S_{photon}= 3.6 \times 10^{45}~k_{B}
.\end{equation}Hence, the entropy generated by the gas of photons is two million times larger than the loss of entropy due to nuclear fusion. The global entropy of the Sun varies by
\begin{equation}
\Delta S_{global} = \Delta S_{fusion} + \Delta S_{photon} = -1.6 \times 10^{39} k_B +  3.6 \times 10^{45} k_B \sim 3.6 \times 10^{45} k_B
\label{EQ:sunS}
.\end{equation}

The loss of entropy due to nuclear fusion is negligible compared to the entropy generated by the photons emitted by the Sun, or any star for that matter. As a result, the global entropy produced by a star can be approximated by the statistical entropy associated with the number of photons produced per unit time.

\section{Comparison of the specific entropy generated by a star and a living body}
\label{secC}%
For the sake of simplicity, we compute here the specific entropy generated by a human body and compare it to that of the Sun. This is equivalent to determining the entropy associated with a gas of photons radiated by a human body and the Sun using Eq.~(\ref{EQ:Sphotons}) (i.e., $S_{photons} =~3.6~k_{B}~N_{photons}$).

As discussed in Appendix~\ref{secB}, the entropy generated by the Sun per unit time is $S_{\odot}$=$3.6 \times 10^{45}~k_{B}$ (Eq.~\ref{EQ:sunS}). Thus the specific entropy (entropy per unit mass) produced by the Sun ($M_{\odot}$=2$\times$10$^{30}$ kg) per second is 

\begin{equation}
S_{\odot}^{spec}= 1.8 \times 10^{15}~k_{B}~~[J~K^{-1}~kg^{-1}~s^{-1}]
.\end{equation}

A human body of 60 kg consumes about 2,500 calories per day, most of which is radiated. Therefore, a human being radiates a power of about 120 W, i.e. 2 W kg$^{-1}$. 
This approximate figure can be estimated otherwise directly from the temperature of a human body.
The internal temperature of the human body is 37$^{\circ}$ C, but the skin temperature of a clothed human is about $T_{body}$ =30$^{\circ}$ C (303 K), at an ambient temperature of $T_{ambient}$ = 20$^{\circ}$C (293 K). The Stefan-Boltzmann law tells us that any physical body brought to a temperature, $T_{body}$, emits light with a power proportional to the temperature to the fourth power. The power per unit area, $P_{body}$, radiated by a human being is equal to the difference between what is radiated and what is absorbed from the surrounding ambient air: 

\begin{equation}
P_{body} = \sigma \times (T_{body}^4[K] - T_{ambient}^4[K]) = 60~[ W~m^{-2} ]
,\end{equation}where $\sigma$ is the Stefan-Boltzmann constant ($\sigma$=5.67$\times$10$^{-8}$ W$ $m$^{-2}$K$^{-4}$).
The surface of an adult human being is about 2 $m^2$ \citep{vangraan1964}. Thus the integrated luminosity of a human body is again equal to 120 W, which is consistent with the previous value (i.e., a specific luminosity of 2 W kg$^{-1}$). At a temperature of $T_{body}$, a human body radiates at a peak wavelength of about 10\,$\mu$m. Thus, after applying Eq.~(\ref{EQ:Sphotons}) (i.e., $S_{photons} =~3.6~k_{B}~N_{photons}$), we find that that a human body generates a specific entropy of

\begin{equation}
S_{human}^{spec}= 3.6 \times 10^{20}~k_{B}~~[J~K^{-1}~kg^{-1}~s^{-1}] = 2 \times 10^5 \times S_{\odot}^{spec}
.\end{equation}As a result, a living animal -- such as a human being -- is 200,000 times more efficient than a star -- such as the Sun -- at producing entropy per unit of mass and time.


\FloatBarrier 
\twocolumn

\FloatBarrier 
\clearpage

\end{appendix}
\end{document}